\documentclass[twocolumn,showpacs,preprintnumbers,amsmath,amssymb]{revtex4}

\usepackage{graphicx}

\begin{document}

\title{Momentum distribution dynamics of a Tonks-Girardeau gas: \\
Bragg reflections of a quantum many-body wavepacket}

\author{R. Pezer and H. Buljan}
\affiliation{Department of Physics, University of Zagreb, 
PP 332, Zagreb, Croatia}

\date{\today}

\begin{abstract}
The dynamics of the momentum distribution and the reduced 
single-particle density matrix (RSPDM) of a Tonks-Girardeau (TG) gas 
is studied in the context of Bragg-reflections of a many-body 
wavepacket. We find strong suppression of a Bragg-reflection peak 
for a dense TG wavepacket; our observation illustrates dependence 
of the momentum distribution on the interactions/wavefunction symmetry. 
The momentum distribution is calculated with a fast algorithm based 
on a formula expressing the RSPDM via a dynamically evolving 
single-particle basis. 
\end{abstract}

\maketitle

The possibility of constraining atomic gases to one-dimensional (1D) 
geometries \cite{OneD,Paredes2004,Kinoshita2004} has lead to experimental 
realizations of exactly solvable 1D models describing interacting 
bose gases \cite{Girardeau1960,Lieb1963}. 
At low temperatures, low linear densities, and strong repulsive effective 
interactions, these 1D atomic gases enter a Tonks-Girardeau (TG) regime 
\cite{Olshanii,Petrov,Dunjko}, which is described by an exactly solvable 
model of 1D bosons with "impenetrable core" repulsive interactions 
\cite{Girardeau1960}.
Two recent experiments achieved the TG regime and observed the 
properties of a TG gas \cite{Kinoshita2004,Paredes2004}. 
One particularly interesting aspect of these 1D systems is 
their nonequilibrium dynamics. A recent experiment studying 
nonequilibrium dynamics of a 1D interacting bose gas 
(including the TG regime) has shown that its momentum distribution 
does not need to relax to thermodynamic equilibrium even after 
numerous collisions \cite{Kinoshita2006}. These experimental 
advances and the possibility of exactly solving the TG model 
\cite{Girardeau1960,Girardeau2000}, motivate us to study 
the momentum distribution of the dynamically evolving TG gas.

The TG model is exactly solvable via Fermi-Bose mapping, 
which relates the TG gas to a system of noninteracting spinless 1D 
fermions \cite{Girardeau1960,Girardeau2000}. 
Many properties of the two systems such as the 
the single-particle (SP) density \cite{Girardeau1960,Girardeau2000} 
or the thermodynamic properties \cite{Das2002} are identical. 
However, quantum correlations contained within the 
reduced single-particle density matrix (RSPDM), or 
the momentum distribution of the TG gas $n_B(k)$, considerably 
differ from those of the ideal Fermi gas $n_F(k)$
\cite{Girardeau2001a,Lenard1964,Girardeau2001,Minguzzi2002,Cazallilla2002,
Papenbrock2003,Forrester2003,Berman2004,Rigol2005,Minguzzi2005,Rigol2006}. 
Although the exact many-body wavefunction describing TG gas 
can be written in compact form \cite{Girardeau1960,Girardeau2000}, 
the calculation of the RSPDM and the momentum distribution is a difficult task 
\cite{Lenard1964,Girardeau2001,Minguzzi2002,Cazallilla2002,
Papenbrock2003,Forrester2003,Berman2004,Rigol2005,Minguzzi2005,Rigol2006}. 
In the stationary case, the RSPDM and $n_B(k)$ were 
studied for a TG gas on the ring \cite{Lenard1964,Forrester2003} 
and in the harmonic confinement 
\cite{Girardeau2001,Minguzzi2002,Papenbrock2003,Forrester2003}. 
In the homogeneous case, the momentum distribution has 
a singularity at $k=0$, $n_B(k)\propto k^{-1/2}$ \cite{Lenard1964}, 
and slowly decaying tails $n_B(k)\propto k^{-4}$ \cite{Minguzzi2002}. 
In both the homogeneous and the harmonic case, the occupation of the leading 
natural orbital (effective SP state) is $\propto \sqrt{N}$ for large $N$
\cite{Forrester2003}. 
An analytic approximation for momentum distribution of a TG gas in a 
box has been made by generalizing the Haldane's harmonic-fluid approach
\cite{Cazallilla2002}.

In the time-dependent case, the RSPDM and momentum distribution of 
the TG gas was studied in a harmonic potential with the time-dependent 
frequency \cite{Minguzzi2005}; dynamics was solved with a 
scaling transformation \cite{Minguzzi2005}. 
Irregular motion, and the dynamics of the momentum distribution, 
was studied numerically for different interaction strengths (up 
to the TG limit) in Ref. \cite{Berman2004}; solutions for 
$N=6$ bosons were presented. 
Several recent studies have addressed the dynamics of hard-core 
bosons (HCB) on the lattice \cite{Rigol2005,Rigol2006}. 
Numerical studies of this model revealed a number of interesting 
results including fermionization of the momentum distribution 
during 1D free expansion \cite{Rigol2005}, and the possibility 
of relaxation of this system to a steady state, which carries 
memory of the initial conditions \cite{Rigol2006}. 
However, the behavior of the {\em discrete} HCB-lattice model is not 
equivalent to the TG bosons in a {\em continuous} potential 
\cite{Cazalilla2004}. 
A feasible numerical study of the RSPDM and related observables 
during motion in a continuous potential $V(x,t)$ demands an efficient 
method for the calculation of the RSPDM, independent of the 
external potential, the state of the system, and which would be operative 
for a larger number of particles.

Here we study dynamics of the momentum distribution, the 
RSPDM, natural orbitals (NOs), their occupancies, and 
Shannon entropy for a TG gas in a continuous 
potential. Our calculation is based on a formula expressing the RSPDM via 
a dynamically evolving SP basis; the method does not 
depend on the external potential, the state of the system, and it is 
operative for a larger number of particles. 
The method is employed in studying Bragg reflections of 
a TG many-body wavepacket in periodic potentials. 
A comparison of the TG bosonic ($n_B$) and noninteractiong fermionic 
($n_F$) momentum distributions illustrates the influence 
of interactions/wavefunction symmetry on this observable. 
The momentum distribution of the ideal fermi gas 
displays a beating peak at the edge of the Brillouin zone. 
In contrast, such a Bragg-reflection peak is completely 
absent for a dense TG wavepacket. 
As the TG wavepacket reflects from the potential, it 
undergoes a rapid decrease of coherence, characterized by the increase 
of entropy and decrease of spatial correlations.

{\em The model.-}
We consider dynamics of $N$ indistinguishable bosons in 1D configuration 
space, located in an external potential $V(x,t)$, and interacting via 
impenetrable pointlike interactions \cite{Girardeau1960}. 
The bosonic many-body wavefunction $\psi_B(x_1,\ldots,x_N,t)$ describing 
the state of this system is related to a fermionic wavefunction 
$\psi_F(x_1,\ldots,x_N,t)$, which describes a system of $N$ noninteracting 
spinless 1D fermions:
$\psi_B(x_1,\ldots,x_N,t)=A(x_1,\ldots,x_N)\psi_F(x_1,\ldots,x_N,t)$,
where $A=\Pi_{1\leq i < j\leq N} \mbox{sgn}(x_i-x_j)$ is a 
"unit antisymmetric function"; this is the famous Fermi-Bose mapping
\cite{Girardeau1960}. 
The dynamics of the fermionic wavefunction $\psi_F$ can be 
constructed from the Slater determinant 
$\psi_F(x_1,\ldots,x_N,t)=\sqrt{1/N!} \det [\psi_m(x_j,t)]$, where 
$\psi_m(x,t)$ denote $N$ orthonormal SP 
wavefunctions $\psi_m(x,t)$ obeying

\begin{equation}
i\hbar \frac{\partial \psi_m}{\partial t}=
\left [ -\frac{\hbar^2}{2m} \frac{\partial^2 }{\partial x^2}+
V(x,t) \right ] \psi_m(x,t), \ m=1,\ldots,N.
\label{master}
\end{equation}
The exact many-body wavefunction of the TG system is

\begin{equation}
\psi_B=A(x_1,\ldots,x_N)\sqrt{\frac{1}{N!}}
\det_{m,j=1}^{N} [\psi_m(x_j,t)],
\label{psi_B}
\end{equation} 
i.e., its evolution is constructed after solving 
Eq. (\ref{master}).

The RSPDM of the TG system, 
$\rho_{B}(x,y,t) = N \int \!\! dx_2\ldots dx_N \,
\psi_B(x,x_2,\ldots,x_N,t)^*
\psi_B(y,x_2,\ldots,x_N,t) $, 
furnishes the expectation values of one-particle observables such 
as the position density $\rho_{B}(x,x,t)$, or momentum distribution 
$n_B(k,t) = (2\pi)^{-1}\int \!\! dx dy \, e^{i k(x-y)}\rho_{B}(x,y,t)$ 
\cite{Lenard1964}. 
The natural orbitals $\phi _i(x,t)$ (NOs) of the TG system,
obtained as eigenfunctions of the RSPDM, 
\[
\int \!\! dy\, \rho_{B}(x,y,t) \, \phi _i (y,t) =
\lambda_i(t) \, \phi _i (x,t), \quad i=1,2,\ldots
\]
represent effective SP states, while eigenvalues $\lambda_i(t)$ 
represent their occupancies \cite{Girardeau2001}. 
The SP wavefunctions $\psi_m(x_j,t)$ 
are NOs of the fermionic system, with occupancy unity, because 
the fermionic RSPDM is 
$\rho_{F}(x,y,t)=\sum_{m=1}^{N}\psi_m^*(x,t)\psi_m(y,t)$ \cite{Girardeau2001}. 
The momentum distributions can be expressed via the 
Fourier transform of the NOs, $n_F(k,t)= \sum_{m=1}^{N} |\tilde\psi_m(k,t)|^2$, 
and $n_B(k,t)= \sum_{i=1}^{\infty} \lambda_i(t)|\tilde\phi_i(k,t)|^2$.

{\em The method.-} The RSPDM can be expressed 
in terms of the dynamically evolving SP basis:

\begin{equation}
\rho_{B}(x,y,t)=\sum_{ij}\psi^{*}_{i}(x,t)A_{ij}(x,y,t)\psi_{j}(y,t).
\label{expansion}
\end{equation}
The $N\times N$ matrix ${\mathbf A}(x,y,t)=\{ A_{ij}(x,y,t) \}$ is 

\begin{equation}
{\mathbf A}(x,y,t)= \det {\mathbf P} ({\mathbf P}^{-1})^{T},
\label{formulA}
\end{equation} 
where the entries of the matrix ${\mathbf P}$ are 
$P_{ij}(x,y,t)=\delta_{ij}-2\int_{x}^{y}dx' \psi_{i}^{*}(x',t)\psi_{j}(x',t)$;
we have assumed $x<y$ without loss of generality.

Derivation of formula (\ref{formulA}) as follows. 
Define permutations 
$(k_2 \ldots k_N)=P(1\ldots i-1\ i+1\ldots N)$, 
$(l_2 \ldots l_N)=Q(1\ldots j-1\ j+1\ldots N)$, and their signatures
$\epsilon(P)$ and $\epsilon(Q)$. From the definition of the RSPDM and 
Eq. (\ref{expansion}) it follows that 

\begin{eqnarray}
A_{ij} & = & \frac{(-1)^{i+j}}{(N-1)!}\int \prod_{n=2}^{N} dx_n 
\mbox{sgn}(x-x_n)\mbox{sgn}(y-x_n)  
\nonumber \\ 
& &    
\sum_{P} \epsilon(P)
\psi_{k_2}^{*}(x_2)\ldots 
\psi_{k_N}^{*}(x_N) 
\nonumber \\ 
& &    
\sum_{Q} \epsilon(Q)
\psi_{l_2}(x_2)\ldots
\psi_{l_N}(x_N)
\label{derivation1} \\
& &  
\nonumber \\ 
& = &  \frac{(-1)^{i+j}}{(N-1)!} 
\sum_{P,Q}  \epsilon(P) \epsilon(Q)
\prod_{n=2}^{N}  P_{k_n,l_n}
\label{derivation2} \\
& = & (-1)^{i+j} 
\det {\mathbf P}_{ij},
\label{derivation3}
\end{eqnarray}
where ${\mathbf P}_{ij}$ is a minor of matrix ${\mathbf P}$ obtained
by crossing its $i$th row and $j$th column. 
Equation (\ref{derivation2}) is obtained after rearranging the 
product factors of Eq. (\ref{derivation1}), and formally performing 
the integrations 
$P_{k_n,l_n}=\delta_{k_n,l_n}-2\int_{x}^{y}dx' \psi_{k_n}^{*}(x',t)\psi_{l_n}(x',t)$. 
Eq. (\ref{derivation3}) follows from the definition of a determinant 
\cite{Forrester2003}. Eq. (\ref{formulA}) follows immediately from 
Eq. (\ref{derivation3}) and the formula for the matrix inverse 
via algebraic co-factors. 

\begin{figure}
\centerline{
\mbox{\includegraphics[width=0.45\textwidth]{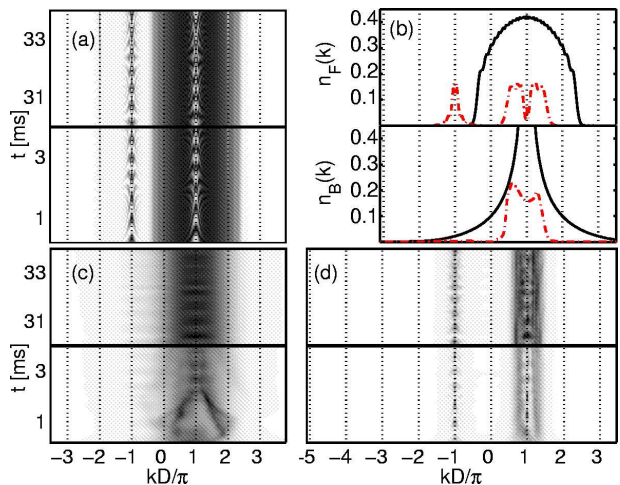}}
}
\caption{ \label{fig1}
(color online) Dynamics of momentum distributions. 
(a) $n_F(k,t)$ in the initial stage of the evolution 
(down), and after long time propagation (up); 
beating at $k=-\pi/D$ is a signature of Bragg reflections. 
(b) Initial momentum distributions $n_F(k,0)$ and $n_B(k,0)$ for $N=25$ 
bosons (solid lines); dot-dashed line depicts 
$\sum_{m=1}^{5}|\tilde \psi(k,t)|^2$ (up), 
and $\sum_{m=1}^{12}\lambda_i(t)|\tilde \phi(k,t)|^2$ (down) 
at $t=34.5$ ms (the area beneath the dot-dashed curves up and down is equal). 
(c) $n_B(k,t)$ for $N=25$ bosons in the initial evolution stage 
(down), and after long time propagation (up); 
the signature of Bragg reflections at $k=-\pi/D$ is absent. 
(d) Dynamics of $n_B(k,t)$ for $N=3$ bosons; 
there is beating at $k=-\pi/D$. 
}
\end{figure}

{\em Dynamics in the periodic potential.-} 
The richness of the dynamics of ultracold bose gases in 
optical lattices \cite{Oberthaler2006} motivate us to 
numerically (exactly) study the evolution of a quantum many-body 
wavepacket in a continuous periodic potential $V_p(x)=V_p(x+D)$ 
(also referred to as the lattice); 
periodic boundary conditions are assumed, i.e., dynamics occurs 
on a ring of length $L=n_s D$. The gas (wavepacket) is initially 
localized within a region significantly smaller than $L$, 
and it is given a certain amount of momentum. 
During dynamics, the many-body wavepacket will disperse on the ring. 
The dynamics of the TG momentum distribution $n_B$ 
is affected by the exchange of the momentum between the lattice 
and the gas, the many-body interactions, and the bosonic symmetry 
of the wavefunction. On the other side, the related 
fermionic momentum distribution $n_F$ is affected by the lattice 
and the Pauli exclusion principle. 
We find it illustrative to compare time-evolution of the 
two momentum distributions, as it illustrates 
the influence of the interactions/wavefunction symmetry on 
this observable.

In our numerical simulations we consider motion of $^{87}$Rb atoms 
in the potential $V_p(x)=V_0 \cos^2(\pi x/D)$, where  
$D=391.5\ \mu$m, and $V_0=11.9$ eV unless specified otherwise; 
$n_s=52$. 
For concreteness, we assume that the 
SP wavefunctions describing the wavepacket at $t=0$ are 
$\psi_m(x,0)=u_m(x)e^{ik'x},\ m=1,\ldots,N$, where $u_m$ is the $m$th SP 
eigenstate of the harmonic potential $V_{h}(x)=m\omega^2/2$, 
$\omega=2\pi\ 316$Hz; such a many-body wavepacket corresponds to 
a ground state of the gas in harmonic confinement, with a momentum 
$k'$ per particle imparted to the wavepacket. 
The initial expectation value of the SP 
momentum $k'=\int \!\! k \, n_B(k,0) dk$ is chosen to be exactly at 
the edge of the Brillouin zone $k'=\pi/D$. 
Although such an excitation is non-trivial to prepare, 
current high level of experimental techniques 
\cite{OneD,Kinoshita2004,Paredes2004,Kinoshita2006} 
strongly suggests that it is more than just a theoretical curiosity.

It should be emphasized that the expectation value of the SP
momentum is identical (at all times) for TG bosons and noninteracting 
fermions, 
$\langle k \rangle_B =\int \!\! dk \, k \, n_B(k,t) 
=\int \!\! dk \, k \, n_F(k,t) =\langle k \rangle_F$. 
Nevertheless, their momentum distributions show remarkable differences. 
Figure \ref{fig1}(a) shows $n_F$ in the initial stage of the evolution, and 
after long-time propagation (when the gas is already well-dispersed over 
the ring). A sharp peak beating up-down at the edge of the 
1st Brillouin zone $k=-\pi/D$ arises from Bragg reflections. 
The fermionic momentum distribution is $n_F(k,t)= \sum_{m=1}^{N} |\tilde\psi_m(k,t)|^2$; 
a few of the SP spectra $|\tilde\psi_m(k,t)|^2$ are initially 
overlapping the edge of the Brillouin zone at $\pi/D$; as the dynamics of 
$\tilde\psi_m(k,t)$ are uncoupled, the spectra $|\tilde\psi_m(k,t)|^2$ of 
those NOs display a beating Bragg-reflection peak at $-\pi/D$
[see dot-dashed curve in Fig. \ref{fig1}(b) up], 
which is reflected onto $n_F(k,t)$.

\begin{figure}
\centerline{
\mbox{\includegraphics[width=0.5\textwidth]{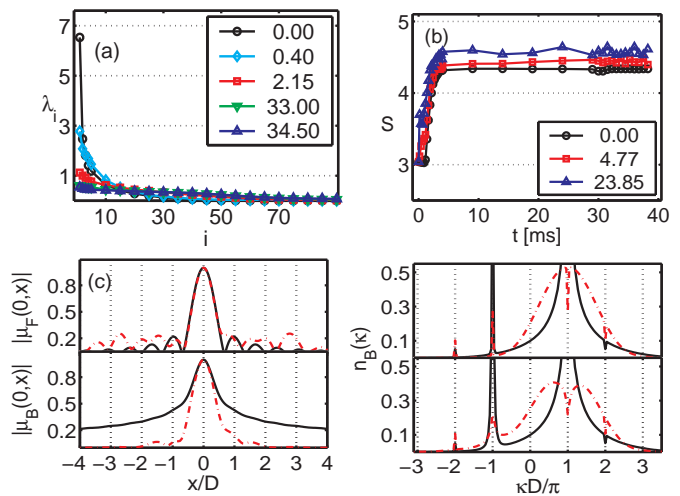}}
}
\caption{
(color online) Dynamics of bosonic NO occupations $\lambda_i$ and entropy $S$. 
(a) $\lambda_i(t)$ at times $t=0,0.40,2.15,33,34.5$ ms;
lattice depth is $V_0=11.9$ eV. (b) The entropy $S(t)$ for 
three different lattice depths $V_0=0,4.77,23.85$ eV.
(c) $|\rho_F(0,x,t)|$ and $|\rho_B(0,x,t)|$ 
at $t=0$ (solid line) and $t=34.5$ (dot-dashed line); $V_0=11.9$ eV. 
(d) The Bloch-wavevector distribution for $25$ bosons 
at $t=0$, and $t=34.5$ ms; upper picture (lower picture) corresponds
to lattice depth $V_0=2.36$ eV ($V_0=11.9$ eV), respectively. 
See text for details. 
}
\label{fig2}
\end{figure}

The bosonic momentum distribution $n_B$ at $t=0$ is shown 
in Fig. \ref{fig1}(b). The sharp peak 
of $n_B(k,0)$ is located exactly at the edge of the Brillouin zone, 
and it is much sharper than the peak of $n_F(k,0)$. 
From this one may erroneously conclude that there would be a sharp 
beating peak originating from Bragg reflections at $-\pi/D$. 
However, this signature of Bragg reflections is {\em absent}. 
This is illustrated in Fig. \ref{fig1}(c), which shows a contour plot
of the bosonic momentum distribution $n_B(k,t)$ for $N=25$ bosons propagating 
in the potential $V_p$. 
The signature is absent both at the beginning of the 
motion, when the wavepacket is still localized, 
and after it spreads over the ring. In the long tome propagation $n_B$ 
collectively oscillates due to the momentum-exchange with the 
lattice ($\langle k \rangle_B =\langle k \rangle_F$), but 
the changes in its shape are small. 
Our simulation clearly depicts that when the momentum 
is being transferred by the lattice to the TG gas, it 
redistributes among bosons; 
this leads to a smooth distribution without a beating Bragg-reflection peak. 
Unlike the fermionic NOs, the low-order bosonic NOs do not display 
Bragg-reflection peaks due to strong (nonlinear) coupling arising 
from interactions [see dot-dashed curve in Fig. \ref{fig1}(b) down].

However, Bragg-reflection peaks can be obtained for a smaller density of the TG gas. 
This is illustrated in Figure \ref{fig1}(d) showing an identical numerical 
simulation but with $N=3$ bosons. In this case, as bosons disperse
on the ring, their density is sufficiently low, leading to the 
fermionization of the bosonic momentum distribution discussed in 
Ref. \cite{Rigol2005}; consequently one observes a beating 
peak at $k=-\pi/D$.

Further insight into Bragg reflections of the TG many-body wavepacket
follows from the behavior of the bosonic NOs $\phi_i$ and their 
occupancies $\lambda_i$. Initially ($t=0$), a few of the 
leading NO occupancies are fairly large [see Fig. \ref{fig2}(a)], 
which is characteristic for a cold Bose gas. However, 
when the evolution begins, the low order 
$\lambda_i$s rapidly decrease, while the number of NOs 
with non-negligible occupations increase [Fig. \ref{fig2}(a)]. 
Figure \ref{fig2}(b) illustrates the time-evolution of the Shannon entropy 
$S(t)=-\sum_i p_i \log p_i$, where $p_i(t)=\lambda_i(t)/N$, 
for different lattice depths $V_0$; the entropy $S$ increases faster, 
and saturates at a higher value for a deeper lattice. 
Figure \ref{fig2}(c) shows bosonic (fermionic) quantum correlations 
$\mu_B(x,x',t)=\rho_B(x,x',t)/\sqrt{\rho_B(x,x,t)\rho_B(x',x',t)}$, 
[$\mu_F(x,x',t)$, respectively] at $t=0$ and $t=34.5$ ms; 
in contrast to $\mu_B$, the correlations of noninteracting fermions 
are not considerably changed during evolution. 
Figures \ref{fig2}(a)-(c) clearly illustrate the dynamical loss of coherence 
of the TG wavepacket, which is more rapid for a deeper lattice. 
This results from the interplay of 
the many-body interactions and scattering from the lattice. 
Namely, interactions couple bosonic NOs thereby providing 
a mechanism for the time-change of their occupancies. 
For a deeper lattice, the initial wavepacket effectively excites 
a larger number of system's eigenstates; this is illustrated in 
Fig. \ref{fig2}(d) which shows the diagonal of the RSPDM represented 
in the Bloch-wave basis (extended Brillouin-zone scheme) 
for two different lattice depths. For a deeper lattice, the dynamics 
effectively involves a larger number of frequencies (i.e., energies), 
and it is more irregular.

Before closing, we note that the dynamics of the TG gas 
is related to the paraxial propagation of 
partially-incoherent light (PIL) beams in linear 1D photonic structures  
\cite{Buljan2006}. Furthermore, the behavior of partially-condensed 
weakly-interacting Bose gases is similar to PIL 
in noninstantaneous nonlinear media \cite{Buljan2005}. 
These facts motivate us to explore the recently observed phenomena 
with incoherent light in photonic lattices \cite{OrenRPLS}, 
within the context of quantum-dynamics of interacting bosons.

In conclusion, we have studied dynamics of the momentum distribution, 
RSPDM correlations, natural orbitals and their occupancies, 
and the entropy of the TG gas out of equilibrium. 
We analyzed Bragg reflections of the TG many-body wavepacket
and found that their signature (observed as a beating 
resonant peak in the momentum distribution of the corresponding noninteracting 
fermionic gas) may be considerably suppressed by the TG many-body interactions. 
We have employed a fast numerical method, applicable for versatile 
{\em continuous} potentials, and operative for larger number of particles. 
Our results open the way for further studies of the RSPDM and 
related observables of the TG gas, both in the static and time-dependent 
cases.

\end{document}